\documentclass[conference,letterpaper,final,twocolumn]{IEEEtran}

\usepackage{amsmath}
\usepackage{amsfonts}
\usepackage{graphicx}

\def\vect#1{{\mathbf #1}}
\def\bc{\vect c}
\def\bd{\vect d}
\def\bu{\vect u}
\def\bv{\vect v}
\def\bx{\vect x}
\def\by{\vect y}
\def\bdh{\hat{\vect d}}
\def\bG{\vect G}
\def\bP{\vect P}
\def\bT{\vect T}
\def\bU{\vect U}
\def\bY{\vect Y}
\def\bigoh{\mathcal O}
\def\calA{\mathcal A}

\newtheorem{theorem}{Theorem}

\title{From Sequential Decoding to Channel Polarization and Back Again}
\author{\IEEEauthorblockN{Erdal Ar{\i}kan}
\IEEEauthorblockA{Department of Electrical and Electronics Engineering\\
	Bilkent University, Ankara, 06800, Turkey}}

\begin{document}
\maketitle

\begin{abstract}
This note is a written and extended version of the Shannon Lecture I gave at 2019 International Symposium on Information Theory. It gives an account of the original ideas that motivated the development of polar coding and discusses some new ideas for exploiting channel polarization more effectively in order to improve the performance of polar codes.
\end{abstract}

\section{Introduction}

\begin{figure}[htb]
\begin{center}
\resizebox{!}{1cm}{
\includegraphics{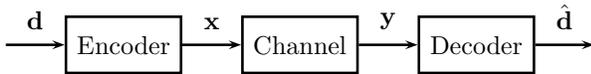}
}
\normalsize
\caption{Channel coding system.}
\label{fig:system}
\end{center}
\end{figure}

We begin with the usual setup for the channel coding problem, as shown in Fig.~\ref{fig:system}.
A message source produces a source word $\bd = (d_1,\ldots,d_K)$ uniformly at random over all possible source words of length $K$ over a finite set, the source word $\bd$ is encoded into a codeword $\bx=(x_1,\ldots,x_N)$, 
the codeword $\bx$ is transmitted over a channel, the channel produces an output word $\by=(y_1,\ldots,y_N)$, and a decoder processes $\by$ to produce an estimate $\bdh = (\hat{d}_1,\ldots,\hat{d}_K)$
of the source word $\bd$. 
The performance metrics for the system are the probability of frame error $P_e =\Pr(\bdh\neq \bd)$, the code rate $R=K/N$, and the complexity of implementation of the encoder and decoder.

\begin{figure}[htb]
\begin{center}
\resizebox{!}{3.7cm}{
\includegraphics{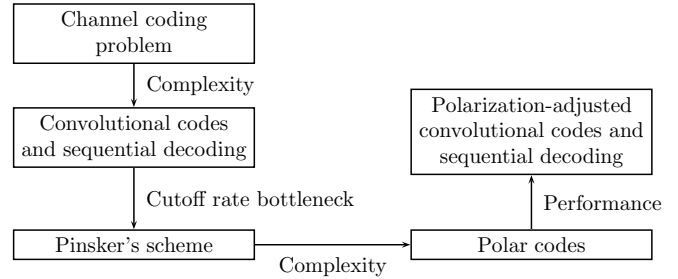}
}
\normalsize
\caption{Order of main topics discussed in the note.}
\label{fig:outline}
\end{center}
\end{figure}

Shannon \cite{shannon_mathematical_1948} proved that for a broad class of channels, there exists a channel parameter $C$, called capacity, such that arbitrarily reliable transmission (small $P_e$) is attainable at any given rate $R$ if $R<C$ (and unattainable if $R>C$). 
Shannon's theorem settled the question about the trade-off between the rate ($R$) and reliability ($P_e$) in a communication system. However, the random-coding analysis Shannon used to prove the attainability  part of his theorem left out complexity issues. Below, we present a track of ideas, as shown in Fig.~\ref{fig:outline}, for constructing practically implementable codes that meet Shannon's capacity bound while providing reliable communication. 

For the rest of the note, we restrict attention to binary-input memoryless channels (BMCs). 
By convention, the channel input alphabet will be $\{0,1\}$, the channel output alphabet will be arbitrary, and the channel transition probabilities will be denoted by $W(y|x)$.
We will also assume that the source alphabet is binary so that $\bd \in \{0,1\}^K$.

Two channel parameters of primary interest will be the symmetric versions of channel capacity and cutoff rate, 
which are defined respectively as
\begin{equation}\label{eq:capacity}
C(W) = \sum_y \sum_{x\in \{0,1\}} \frac{1}{2}W(y|x) \log_2 \frac{W(y|x)}{\frac{1}{2} W(y|0)+\frac{1}{2} W(y|1)}
\end{equation}
and
\begin{equation}\label{eq:cutoff}
R_{0}(W) = 1- \log_2\bigg(1+\sum_y \sqrt{W(y|0)W(y|1)}\bigg).
\end{equation}
If the BMC under consideration happens to have some symmetry properties as defined in \cite[p.~94]{r._g._gallager_information_1968}, then the symmetric capacity and symmetric cutoff rate 
coincide with their true versions (which are obtained by an optimization over all possible distributions on the channel input alphabet).
For our purposes, the symmetric versions of the capacity and cutoff rate are more relevant than their true versions since throughout this note we will be considering {\sl linear} codes. 
Linear codes are constrained to use the channel input symbols 0 and 1 with equal frequency so they can at best achieve the symmetric capacity and symmetric cutoff rate.
For brevity, in the rest of the note, we will omit the qualifier ``symmetric'' when referring to $C(W)$ and $R_0(W)$; the reader should remember that all such references are actually to the symmetric versions of these parameters
as defined by \eqref{eq:capacity} and \eqref{eq:cutoff}. 

A third channel parameter that will be useful in the following is the Bhattacharyya parameter defined as 
\begin{equation}\label{eq:Bhattacharyya}
Z(W)=\sum_y \sqrt{W(y|0)W(y|1)}.
\end{equation}
We note the relation $R_0(W) = 1 - \log_2\big[1+Z(W)\big]$, which will be important in the sequel.

\section{Convolutional codes and sequential decoding}

\begin{figure}[htb]
\begin{center}
\resizebox{!}{4cm}{
\includegraphics{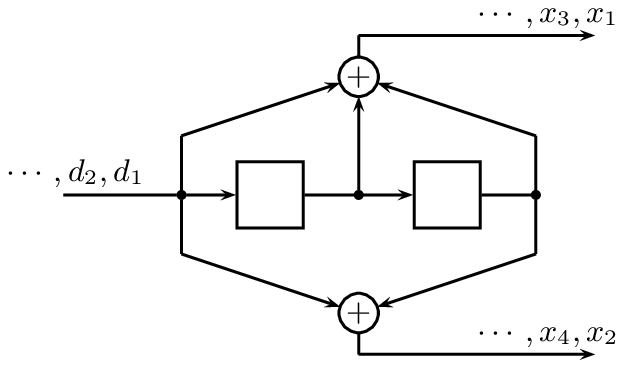}
}
\normalsize
\caption{Example of a convolutional code.}
\label{fig:ConvCode}
\end{center}
\end{figure}

Convolutional codes are a class of linear codes introduced by Elias \cite{peter_elias_coding_1955} with an encoder mapping of the form
$\bx=\bd\bG$ 
where the generator matrix $\bG$ has a special structure that corresponds to a convolution operation. An example of a convolutional code is one with the generator matrix
\setcounter{MaxMatrixCols}{15}
\begin{equation*}
\bG = \begin{bmatrix} 
1 & 1 & 1 & 0 & 1 & 1 &   &   &   &   &   & & & \\
  &   & 1 & 1 & 1 & 0 & 1 & 1 &   &   &   & & & \\
  &   &   &   & 1 & 1 & 1 & 0 & 1 & 1 &   & & & \\  
  &   &   &   &   &   & 1 & 1 & 1 & 0 & 1  & 1 & \\  
  &   &   &   &   &   &   &   &  & & &   &   &  
\end{bmatrix},
\end{equation*}
for which the encoding operation can be implemented using the convolution circuit in Fig.~\ref{fig:ConvCode}.

\begin{figure}[thb]
\begin{center}
\resizebox{!}{10cm}{
\includegraphics{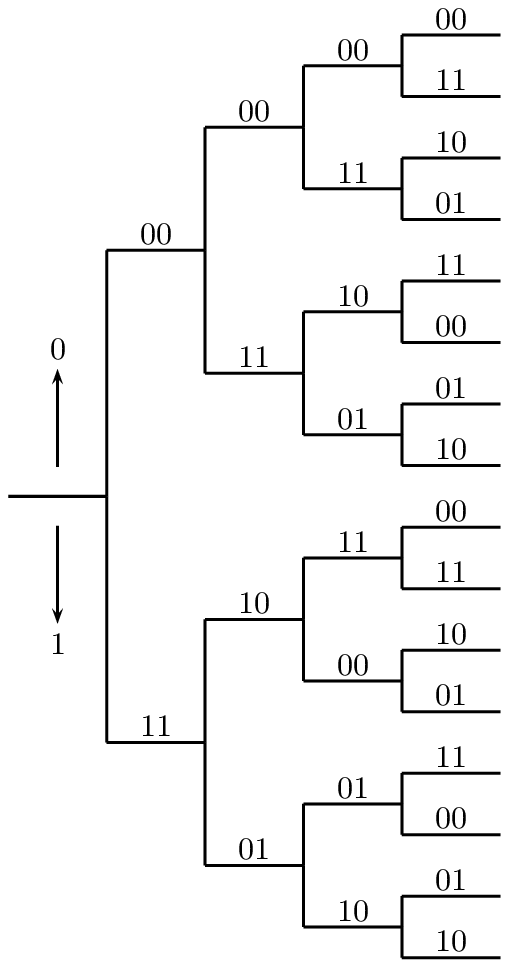}
}
\normalsize
\caption{Tree representation of a convolutional code.}
\label{fig:TreeCode}
\end{center}
\end{figure}

The codewords of a convolutional code can be represented in the form of a tree.
For example, the first four levels of the tree corresponding to the convolutional code of Fig.~\ref{fig:ConvCode} are shown in Fig.~\ref{fig:TreeCode}.  
Each source word $\bd=(d_1,\ldots,d_K)$ defines a path through the code tree (take the upper branch if $d_i$ is 0, the lower branch otherwise). 
Branches along a path are labeled with the codeword symbols corresponding to that path. 

The tree representation of a convolutional code turns the decoding problem into a tree search problem. 
One of the paths through the tree is the correct path and all other paths are incorrect paths. Exhaustive search of the tree for the correct path corresponds to optimum decoding but is too complex to implement.
There is need for low-complexity tree search heuristics that can be used as decoders. 
A reasonable choice is a depth-first search heuristic. 
Sequential decoding is a depth-first search heuristic developed by Wozencraft \cite{wozencraft_sequential_1957} for decoding arbitrary tree codes.

The computational complexity in sequential decoding (the number of steps it takes to complete decoding) is a random variable whose statistical properties (mean, variance, distribution)
depend on the code rate and the channel characteristics. 
Sequential decoding achieves the capacity $C(W)$ of any given BMC $W$ if no limit is placed on its search complexity. 
However, the average complexity in sequential decoding becomes prohibitive for practical purposes if the code rate is above the cutoff rate $R_0(W)$. 
More precisely, at rates $R > R_0(W)$, the average complexity of decoding the first $nR$ source bits {\sl correctly} is lower-bounded roughly as $2^{n[R-R_0(W)]}$, 
while at rates $R< R_0(W)$ virtually error free communication is possible at constant average complexity per decoded bit. 
Detailed accounts of the sequential decoding algorithm and its complexity may be found in \cite[pp.~263-286]{r._g._gallager_information_1968} and \cite[pp.~425-476]{j._m._wozencraft_principles_1965}.

My interest in sequential decoding goes back to 1983 when I was a doctoral student at M.I.T. and my thesis supervisor Bob Gallager asked me to look at sequential decoding for multiaccess channels. 
This subject became my PhD thesis \cite{arikan_sequential_1985}.
Multiaccess communications was an emerging subject and sequential decoding was a good starting point for assessing the practical viability of coding for multiaccess channels (see \cite{gallager_perspective_1985} for the broader context of this problem). Historically, sequential decoding had been a method of choice briefly (used in space communications (Pioneer 9, 1968)) before being superseded by Viterbi decoding in the 1970s.
Despite having fallen out of favor, sequential decoding was still an interesting subject with rich connections to information theory and error exponents. 
In studying sequential decoding, I came across two fascinating papers by Pinsker \cite{pinsker_complexity_1965}
and Massey \cite{massey_capacity_1981}. 
These papers showed how to ``boost'' the cutoff rate of sequential decoding in a sense described below. 
An extended discussion of both papers as they relate to my later work on polar coding can be found in \cite{arikan_origin_2016}. 
In the following, I will focus mainly on \cite{pinsker_complexity_1965} because of its general nature. 
However, before proceeding to \cite{pinsker_complexity_1965}, I will review \cite{massey_capacity_1981} since it contains some of the essential ideas in this note in a very simple setting.

\section{Massey's example}\label{Sect:Massey}
Let $M=2^m$ for some integer $m\ge 2$, and consider an $M$'ary erasure channel (MEC) with input alphabet ${\mathcal X}= \{0,1,\ldots,2^m-1\}$, 
output alphabet ${\mathcal Y}={\mathcal X}\cup \{?\}$ (where $?$ is an erasure symbol), and transition probabilities $W(y|x)$ such that, when $x\in {\mathcal X}$ is sent, 
the channel output $y$ has two possible values, $y=x$ and $y=?$, which it takes with conditional probabilities $W(x|x)=1-\epsilon$ and $W(?|x)=\epsilon$.  
The capacity and cutoff rate of the MEC are readily calculated as $C(m) = m(1-\epsilon)$ and $R_0(m) = m -\log_2\big(1+(2^m-1)\epsilon\big)$.

Massey observed that the MEC can be split into $m$ binary erasure channels (BECs) by relabeling its inputs and outputs with vectors of length $m$.
A specific labeling that achieves this is as follows. 
Each input symbol $x\in {\mathcal X}$ is relabeled with its binary representation $(x_1,\ldots,x_m)\in \{0,1\}^m$ so that $x=\sum_{i=1}^m x_i2^{m-i}$. 
Each output symbol $y\in {\mathcal Y}$ is relabeled with a vector $(y_1,\ldots,y_m)$ which equals the binary representation of $y$ if $y\in {\mathcal X}$ and equals $(?,\ldots,?)$ if $y=?$.
With this relabeling, a single transmission event $\{(x_1,\ldots,x_m) \to (y_1,\ldots,y_m)\}$ across the MEC can be thought of as a collection of $m$ transmission events $\{x_i\to y_i\}$ across the coordinate channels.
An erasure event in the MEC causes an erasure event in all coordinate channels; if there is no erasure in the MEC, there is no erasure in any of the coordinate channels.
Each coordinate channel is a BEC with erasure probability $\epsilon$. 
The coordinate channels are fully correlated in the sense that when an erasure occurs in one of them, an erasure occurs in all of them.

The capacity and cutoff rate of the BECs are given by $C(1) = 1-\epsilon$ and $R_0(1) = 1-\log_2(1+\epsilon)$.
It can be verified readily that $C(m)=mC(1)$ (capacity is conserved), while $R_0(m) \le  m R_0(1)$ with strict inequality unless $\epsilon$ equals 0 or 1.
Thus, splitting the MEC does not cause a degradation in channel capacity but ``improves'' or ``boosts'' the cutoff rate.
This example shows that one may break the cutoff rate barrier for the MEC by employing a separate convolutional encoder -- sequential decoder pair on each coordinate BEC.
The reader is advised to see \cite{gallager_perspective_1985} for an alternative look at this important example from the perspective of multiaccess channels.
To learn about the communications engineering context in which Massey's example arose, we refer to \cite{massey_capacity_1981}.

Massey's example provides a basis for understanding the more complex schemes presented below. 
These more complex schemes begin with independent copies of a binary-input channel (raw channels), build up a large channel (akin to the MEC) through some channel combining operations, and then split the large channel back to a set of correlated binary-input channels (synthesized channels). One speaks of a ``boosting'' of the cutoff rate if the sum of the cutoff rates of the synthesized channels is larger than the sum of the cutoff rates of the raw channels.

\section{Pinsker's scheme}

\begin{figure}
\begin{center}
\resizebox{!}{6cm}{
\includegraphics{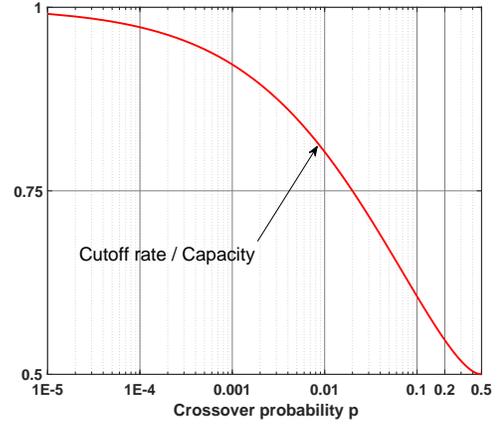}
}
\normalsize
\caption{Ratio of cutoff rate to capacity for the BSC.}
\label{fig:BSC}
\end{center}
\end{figure}

Pinsker \cite{pinsker_complexity_1965} observed that, for the binary symmetric channel (BSC) with crossover probability $p$ (a BMC with output alphabet $\{0,1\}$ and $W(1|0)=W(0|1)=p$), the ratio of the cutoff rate to capacity approaches 1 as $p$ goes to 0, 
$$
\frac{R_0}{C} =\frac{1-\log_2\big[1+2\sqrt{p(1-p)}\,\big]}{1+p\log_2(p)+(1-p)\log_2(1-p)} \rightarrow 1 \quad \text{as $p\to 0$},
$$
as illustrated in Fig.~\ref{fig:BSC}. 
Pinsker combined this observation with Elias' product coding idea \cite{elias_error-free_1954} to construct a coding scheme that boosted the cutoff rate to capacity. 

\begin{figure*}[thb]
\begin{center}
\resizebox{!}{5cm}{
\includegraphics{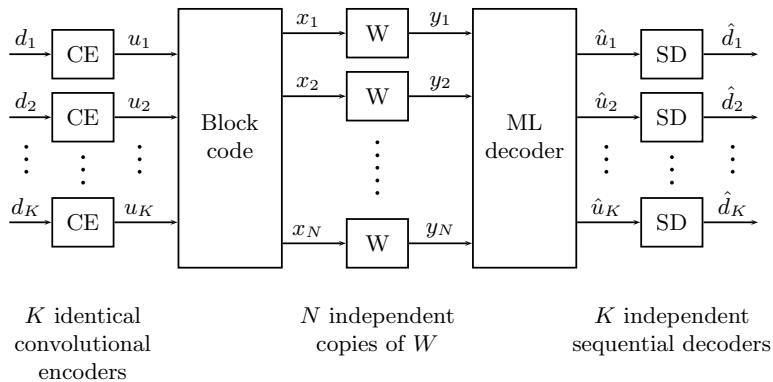}
}
\normalsize
\caption{Pinsker's scheme.}
\label{fig:Pinsker1}
\end{center}
\end{figure*}

\begin{figure}[thb]
\begin{center}
\resizebox{!}{4cm}{
\includegraphics{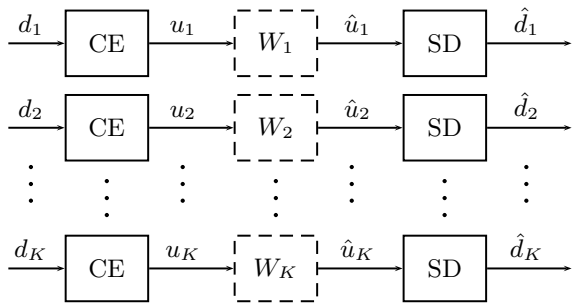}
}
\normalsize
\caption{Bit-channels created by Pinsker's scheme.}
\label{fig:Pinsker2}
\end{center}
\end{figure}

Pinsker's scheme, as shown in Fig.~\ref{fig:Pinsker1}, uses an inner block code and $K$ identical outer convolutional codes. 
Each round of operation of the inner block code comprises the encoder for the inner block code receiving one bit from the output of each outer convolutional encoder (for a total of $K$ bits) and encoding them into an inner code block of length $N$ bits. 
The inner code block is then sent over a BMC $W$ by $N$ uses of $W$.
Since successive bits at the output of each outer convolutional encoder are carried in separate inner code blocks, they suffer i.i.d. error events. So, each outer convolutional code sees a {\sl memoryless} bit-channel, as depicted in Fig.~\ref{fig:Pinsker2}. 
We denote by $W_i:U_i\to \hat{U}_i$ the (virtual) BMC that connects the $i$th convolutional encoder to the $i$th sequential decoder.\footnote{We use capital letters $U_i$ and $\hat{U}_i$ to denote the random variables corresponding to $u_i$ and $\hat{u}_i$. This convention of using capital letters to denote random variables is followed throughout.}

To show that this scheme is capable of boosting the cutoff rate arbitrarily close to channel capacity, we may fix the rate $K/N$ of the inner block code as $(1-\delta)C(W)$ for some constant $0<\delta <1$ and consider increasing the block length $N$ and choosing a good enough inner block code so as to ensure that the bit-channels $W_1,\ldots,W_K$ become near-perfect with $R_0(W_i) >1-\epsilon$ for each $i$, where $\epsilon>0$ is a second constant independent of $N$ and $i$. 
This ensures that each outer convolutional code can operate at a rate $1-\epsilon$ and still be decoded by a sequential decoder at an average complexity bounded by a third constant, where the third constant depends on $\delta$ and $\epsilon$ but not on $N$.
The overall rate for this scheme is $K(1-\epsilon)/N = (1-\delta)(1-\epsilon)C(W)$, which can be made arbitrarily close to $C(W)$ by choosing $\delta$ and $\epsilon$ sufficiently small.
In Pinsker's words, his scheme shows that ``[f]or a very general class of channels operating below capacity it is possible to construct a code in such a way that the number of operations required for decoding is less than some constant that is independent of the error probability''.

Pinsker's result complements Shannon's result by showing that, at any fixed rate $R$ below channel capacity $C(W)$, the average complexity per decoded bit can be kept bounded by a constant while achieving any desired frame error rate $P_e>0$.
Unfortunately, the recipe for choosing a good enough inner block code in Pinsker's scheme is to pick the code at random. 
The non-constructive nature of Pinsker's scheme and the complexity of ML decoding of a randomly chosen block code make Pinsker's scheme impractical.
For our purposes, the takeaway from Pinsker's scheme is the demonstration that there is no ``cutoff rate barrier to sequential decoding'' in a fundamental sense. 
Our next goal will be to find a way of breaking the cutoff rate barrier in a practically implementable manner.

Before we end this section, it is instructive to compare Pinsker's scheme with Massey's example. In Massey's example, a given channel is split into multiple correlated bit-channels.
In Pinsker's scheme, the first step is to synthesize a large channel from a collection of independent bit-channels; the large channel is then split back into a number of dependent bit-channels.
Massey's example appears to be a very special case that cannot be generalized to arbitrary BMCs, while Pinsker's scheme is entirely general.
Massey's example boosts the cutoff rate almost effortlessly but cannot boost it all the way to channel capacity. Pinsker's scheme is much more complex but can boost the cutoff rate to capacity.
Both schemes use multiple sequential decoders.
The use of multiple sequential decoders is a crucial aspect of both schemes.
If a single sequential decoder were used in Pinsker's scheme to decode all $K$ convolutional codes jointly (using a joint tree representation), then a ``data-processing'' theorem by Gallager \cite[pp.~149-150]{r._g._gallager_information_1968} would limit the achievable cutoff rate to $R_0(W)$.
For more on this point, we refer to \cite{arikan_origin_2016}.

\section{Multi-level coding}

\begin{figure*}[thb]
\begin{center}
\resizebox{!}{6cm}{
\includegraphics{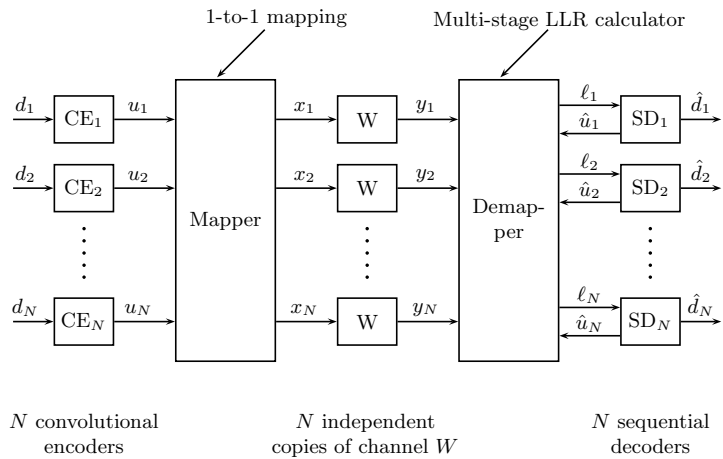}
}
\normalsize
\caption{Multi-level coding}
\label{fig:MLC}
\end{center}
\end{figure*}

In order to reduce the complexity in Pinsker's scheme, in this section, we look at multi-level coding (MLC) with multi-stage decoding (MSD), a scheme due to Imai and Hirakawa \cite{imai_new_1977}.
The MLC/MSD system makes better use of the information available at the receiver and hence it has the potential to boost the cutoff rate at lower complexity.
The particular MLC/MSD system we consider here is shown in Fig.~\ref{fig:MLC}. The mapper in the figure is a one-to-one transformation. The demapper is a 
device that calculates sufficient statistics in the form of log-likelihood ratios (LLRs) and feeds them to a MSD unit.
Each decoder in the MSD chain is able to benefit from the decisions by the previous decoders in the chain. 

\begin{figure}[thb]
\begin{center}
\resizebox{!}{3.5cm}{
\includegraphics{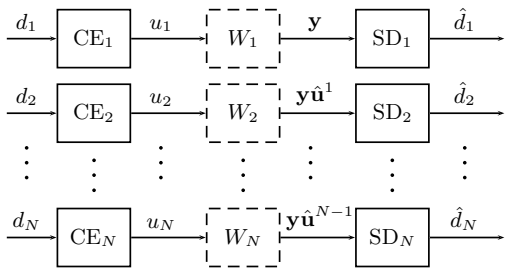}
}
\normalsize
\caption{Bit channels created by MLC/MSD}
\label{fig:MLC2}
\end{center}
\end{figure}

In effect, the MLC/MSD system creates $N$ bit-channels $W_1,\ldots, W_N$, as shown in Fig.~\ref{fig:MLC2}, where the $i$th bit-channel is of the form 
$W_i:U_i\to \bY\hat{\bU}^{i-1}$.
More precisely, $W_i$ is the channel whose input $U_i$ is a bit taken from the output of the $i$th convolutional encoder and whose output $\bY\hat{\bU}^{i-1}$ is the input to the $i$th sequential
decoder in the MSD chain. Here, $\bY=(Y_1,\ldots,Y_N)$ is the entire channel output vector and $\hat{\bU}^{i-1}=(\hat{U}_1,\ldots,\hat{U}_{i-1})$ is the vector of decisions provided by the decoders that precede decoder $i$ in the MSD chain. 

If the MLC/MSD system is configured so that the sequential decoders provide virtually error-free decisions, then the bit-channel $W_i$ takes the form $W_i:U_i\to \bY\bU^{i-1}$
where the decisions fed forward by the previous stages are always correct.
For purposes of deriving polar codes, it suffices to consider only this ideal case with no decision errors.
Hence, from now on, we suppose that $W_i$ has this ideal form.

An important property of the MLC/MSD scheme is the conservation of capacity,
$$
\sum_{i=1}^N C(W_i) = \sum_{i=1}^N I(U_i;\bY\bU^{i-1}) = I(\bU^N;\bY^N) = NC(W),
$$
where the second equality is obtained by writing $I(U_i;\bY\bU^{i-1})=I(U_i;\bY|\bU^{i-1})$ based on the assumption that $U_i$ and $\bU^{i-1}$ are independent and then using the chain rule.

The MLC/MSD scheme conserves capacity at any finite construction size $N$ while Pinsker's scheme conserves capacity only in an asymptotic sense.
Thus MLC/MSD uses information more efficiently and hence may be expected to achieve a given performance at a lower construction size (leading to a lower complexity).

On the other hand, unlike Pinsker's scheme in which the outer convolutional codes are all identical, the natural rate assignment for the MLC/MSD scheme is to set the rate $R_i$ of the $i$th convolutional code to 
a value just below $R_0(W_i)$.
Using convolutional codes at various different rates $\{R_i\}$ as dictated by $\{R_0(W_i)\}$, and decoding them using a chain of sequential decoders 
is a high price to pay for the greater information efficiency of the MLC/MSD scheme. 
Fortunately, this complexity issue regarding outer convolutional codes and sequential decoders is not as severe as it looks thanks to a phenomenon called {\sl channel polarization}.

\begin{theorem}\label{thm:Polarization}
Consider a sequence of MLC/MSD schemes over a BMC $W$, with the $n$th scheme in the sequence having size $N=2^n$ and a mapper of the form
\begin{equation}\label{eq:Pn}
\bP_{n} = \begin{bmatrix} 1 & 0 \\ 1 & 1 \end{bmatrix}^{\otimes n},
\end{equation}
where the exponent ``$\otimes n$'' indicates the $n$th Kronecker power.
Fix $0<\delta< \frac12$. 
As $n$ increases, the idealized bit-channels $\{W_i\}_{i=1}^N$ for the $n$th MLC/MSD scheme polarize in the sense that the fraction of channels with $C(W_i)>1-\delta$ tends to $C(W)$ and the fraction with $C(W_i)< \delta$ tends to $1-C(W)$.
For each bit-channel $W_i$ that polarizes, its cutoff rate $R_o(W_i)$ polarizes to the same point (0 or 1) as its capacity $C(W_i)$.
Furthermore, the mapper and demapper functions can be implemented at complexity $\bigoh(N\log N)$ per mapper block $\bu$. 
\hfill$\diamond$
\end{theorem}

We refer to \cite{arikan_channel_2009} for a proof of this theorem.

The most important aspect of Theorem~\ref{thm:Polarization} is its statement that polarization can be achieved at complexity $\bigoh(\log N)$ per transmitted bit.
In the absence of a complexity constraint, polarization alone is not hard to achieve. A randomly chosen mapper is likely to achieve polarization but is also likely to be too complex to implement.
The recursive structure of the mappers $\{\bP_n\}$ used in Theorem~\ref{thm:Polarization} make it possible to obtain polarization at low complexity.
We will see below that the polarization effect brought about by the transforms $\{\bP_n\}$ is strong enough to simplify the rate assignment $\{R_i\}$ while also maintaining
reliable transmission of source data bits after the MLC/MSD scheme is simplified. 
However, we first wish to illustrate the polarization phenomenon of Theorem~\ref{thm:Polarization} by an example.

\begin{figure}[thb]
\begin{center}
\resizebox{!}{5cm}{%
\includegraphics{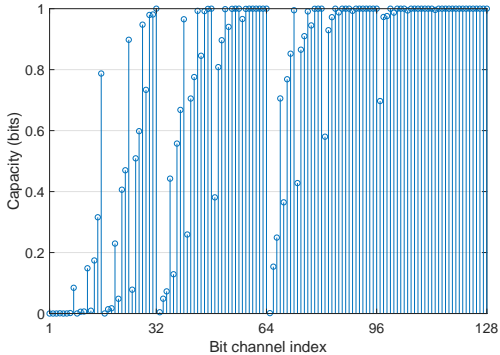}
}
\normalsize
\caption{Channel polarization for BIAWGN channel at 3 dB SNR.}
\label{fig:BIAWGNPol}
\end{center}
\end{figure}

In Fig.~\ref{fig:BIAWGNPol}, we show a plot of $C(W_i)$ v. $i$ for the bit-channels $\{W_i\}$ created by an MLC/MSD construction of size $N=128$ using the transform $\bP_n$ with $n=7$.
The channel in the example is a binary-input additive white Gaussian noise (BIAWGN) channel, which is a channel that receives a binary symbol $x\in\{0,1\}$ as input, 
maps it into a real number $s$ by setting $s=1$ if $x=0$ and $s=-1$ otherwise, 
and generates a channel output $y=s+z$, where $z\sim N(0,\sigma^2)$ is additive Gaussian noise independent of $s$.   
The signal-to-noise ratio (SNR) for the BIAWGN channel is defined as $1/\sigma^2$. 
The SNR in Fig.~\ref{fig:BIAWGNPol} is 3 dB.
The capacity $C(W)$ of the BIAWGN channel $W$ at 3 dB SNR is $0.72$ bits; hence, by Theorem~\ref{thm:Polarization}, we expect that roughly a fraction 0.72 of the capacity terms $C(W_i)$ in Fig.~\ref{fig:BIAWGNPol}
will be near 1. 

\begin{figure}[thb]
\begin{center}
\resizebox{!}{0.7\columnwidth}{
\includegraphics{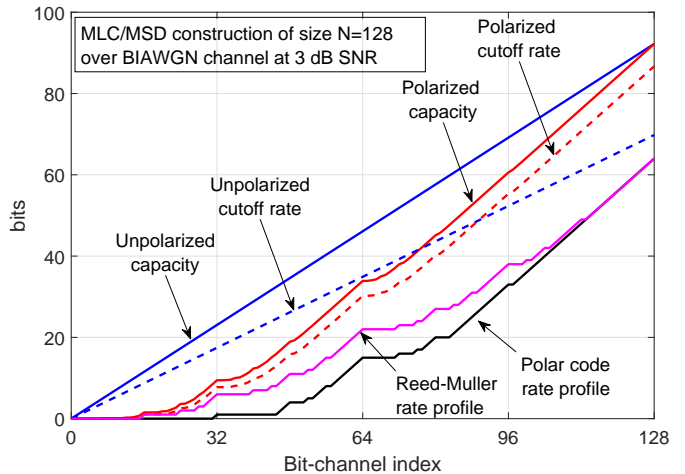}
}
\normalsize
\caption{Capacity and cutoff rate profiles over BIAWGN channel.}
\label{fig:Profile}
\end{center}
\end{figure}

An alternative view of the channel polarization effect in the preceding example is presented in Fig.~\ref{fig:Profile} where cumulative distributions ({\sl profiles}) of various information parameters are plotted 
as a function of an index variable $i$ which takes values from 0 to $N=128$.
The polarized capacity profile is defined as the sequence of cumulatives $\big\{\sum_{j=1}^i C(W_j)\big\}$ indexed by $i$. 
Likewise, the polarized cutoff rate profile is defined as $\big\{\sum_{j=1}^i R_0(W_j)\big\}$, the
unpolarized capacity profile as $\big\{iC(W)\big\}$, and
the unpolarized cutoff rate profile as $\big\{iR_0(W)\big\}$.
By convention, we start each profile at 0 at $i=0$.
The two other curves in the figure (Reed-Muller and polar code rate profiles) will be discussed later.

The unpolarized capacity and cutoff rate profiles in Fig.~\ref{fig:Profile} serve as benchmarks, corresponding to the case where the mapper in the MLC scheme is the identity transform.
The polarized capacity and cutoff rate profiles demonstrate the polarization effect due to the transform $\bP_7$.
The polarized and unpolarized capacity profiles coincide at $i=0$ and $i=N$, but a gap exists between the two for $0< i< N$ due to channel polarization. 
Ideally, the polarized capacity profile would stay zero until $i$ is around $[1-C(W)]N = 35.8$ and then climb with a slope of 1 until  $i=N$. 
A mapper chosen at random is likely to create a near-ideal polarized capacity profile, but the corresponding demapper function is also likely to be too complex. 
By using $\bP_7$ as the mapper, we settle for a non-ideal polarized capacity profile in return for lower implementation complexity.

A beneficial by-product of channel polarization is the boosting of the cutoff rate, which is clearly visible in Fig.~\ref{fig:Profile}. 
The polarized cutoff rate profile has a final value $\sum_{i=1}^N R_0(W_i) = 86.7$ compared to a final value $NR_0(W)=69.8$ for the unpolarized cutoff rate profile. 
Theorem~\ref{thm:Polarization} ensures that, asymptotically as $N$ becomes large, the normalized sum cutoff rate $\frac1N \sum_{i=1}^N R_0(W_i)$ approaches $C(W)$. 
So, the MLC/MSD scheme, equipped with the transforms $\{\bP_n\}$, reproduces Pinsker's result by boosting the cutoff rate to channel capacity, with the important difference that here the mapper and demapper complexity per transmitted source bit is $\bigoh(\log N)$ for a construction of size $N$ (while the similar complexity in Pinsker's scheme is exponential in $N$).

Despite the reduced mapper/demapper complexity, the MLC/MSD scheme (with the transforms $\{\bP_n\}$) is still far from being practical since it calls for using $N$ outer convolutional 
codes at various code rates. 
At this point, we take advantage of the polarization effect and constrain the rates $R_i$ to 0 or 1. 
Such a 0-1 rate assignment in effect eliminates the outer codes.
Setting $R_i=0$ corresponds to fixing the input to the $i$th bit channel $W_i$. 
Setting $R_i=1$ corresponds to sending information in uncoded form over the $i$th bit-channel $W_i$.
In either case, the MSD decisions can be made independently from one mapper block (of length $N$) to the next, eliminating the need for a sequential decoder. 

The 0-1 rate assignment leads to a new type of stand-alone block code, which we will call a {\sl polar code}.
The simplified MSD function under the 0-1 rate assignment will be called {\sl successive cancellation} (SC) decoding.
An important new question that arises is whether polar codes, obtained by such drastic simplification of the MLC/MSD scheme, can provide reliable transmission of source data.
An answer to this question is provided in the next section.

\section{Polar codes}

In this section we will study polar codes as a stand-alone coding scheme. 
For simplicity, we will consider polar coding only for BMCs that are symmetric in the sense defined in \cite{arikan_channel_2009} or \cite[p.~94]{r._g._gallager_information_1968}.
We begin by restating the definition of polar codes without any reference to their origin.
 
A {\sl polar code} is a linear block code characterized by three parameters: a code block-length $N$, a code dimension $K$, and a data index set $\calA$.
The code block-length is constrained to be a power of two, $N=2^n$ for some $n\ge 1$. The code dimension can be any integer in the range $1\le K\le N$.
The data index set $\calA$ is a subset of $\{1,\ldots,N\}$ with size $|\calA|=K$. (This set corresponds to the set of indices $i$ for which $R_i=1$ in the MLC/MSD context.)
A method of choosing $\calA$ will be given below. 
The encoder for a polar code with parameters $(N,K,\calA)$ receives a source word $\bd$ of length $K$ and embeds it in a carrier vector $\bu$ so that $\bu_\calA=\bd$ and $\bu_{\calA^c}={\mathbf 0}$. 
(Here, $\bu_\calA=(u_i:i\in \calA)$ is a subvector of $\bu$ obtained by discarding all coordinates outside $\calA$.) 
Encoding is completed by computing the transform $\bx =\bu \bP_n$, where $\bP_n$ is as defined in \eqref{eq:Pn}.
Henceforth, we will refer to $\bP_n$ as a {\sl polar transform}.

The standard decoding method for polar codes is SC decoding. For details of SC decoding, we refer to \cite{arikan_channel_2009}.
As shown in \cite{arikan_channel_2009}, for a symmetric BMC $W$, the probability of frame error $P_e$ for a polar code under SC decoding is bounded as
\begin{equation}\label{eq:bound}
P_e \le \sum_{i\in \calA} Z(W_i)
\end{equation}
where $Z(W_i)$ is the Bhattacharyya parameter of channel $W_i$. 
From now on, we will assume that the data index set $\calA$ is chosen so as to minimize the bound \eqref{eq:bound} on $P_e$, {\sl i.e.}, 
that $\calA$ is selected as a set of $K$ indices $i$ such that $Z(W_i)$ is among the $K$ smallest numbers in the list $Z(W_1),\ldots,Z(W_N)$.
Since $Z(W_i)=2^{1-R_0(W_i)}-1$, an equivalent rule for constructing a polar code is to select $\calA$ as a set of $K$ indices $i$ such that $R_0(W_i)$ is among the $K$ largest cutoff rates in the list $R_0(W_1),\ldots,R_0(W_N)$.

\begin{theorem}\label{thm:PolarCodes}
A polar code with length $N$, dimension $K$, and rate $R=K/N$ over a symmetric BMC $W$ has the following properties.
\begin{itemize}
\item It can be constructed (the data index set ${\cal A}$ can be determined) in $\bigoh(N\text{poly}(\log N))$ steps \cite{mori_performance_2009}, \cite{pedarsani_construction_2011}, \cite{tal_how_2013}.
\item It can be encoded and SC-decoded in $\bigoh(N\log N)$ steps \cite{arikan_channel_2009}.
\item Its frame error rate $P_e$ under SC decoding is bounded as $\bigoh(e^{-N^{0.499}})$ for any fixed rate $R< C(W)$ \cite{arikan_rate_2009}.
\end{itemize}
\hfill$\diamond$
\end{theorem}

In summary, polar coding achieves the capacity of symmetric BMCs with low-complexity encoding, decoding, and construction methods.
For a precise discussion of the novelty of polar codes as a capacity-achieving code construction, we refer to \cite{guruswami_polar_2015}.

\begin{figure}[thb]
\begin{center}
\resizebox{!}{0.6\columnwidth}{
\includegraphics{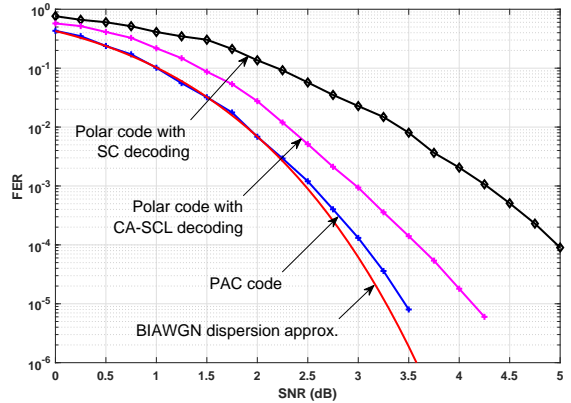}
}
\normalsize
\caption{Performance curves over the BIAWGN channel.}
\label{fig:Performance}
\end{center}
\end{figure}

The performance of polar codes is far from optimal.
Fig.~\ref{fig:Performance} illustrates the frame error rate (FER) $P_e$ under SC decoding of a polar code with block-length $N=128$ and rate $R=1/2$ over a BIAWGN channel with the SNR ranging from 0 to 5 dB. 
This and other FER curves in Fig.~\ref{fig:Performance} have been obtained by computer simulation.
Also shown in Fig.~\ref{fig:Performance} is the BIAWGN dispersion approximation \cite{polyanskiy_channel_2010} at block-length $N=128$ and rate $R=1/2$, which is an estimate of the average ML-decoding performance
over the BIAWGN channel of a code chosen uniformly at random from the ensemble of all possible binary codes of block-length $N=128$ and rate $R=1/2$. 

The weak performance of polar codes is due in part to the suboptimality of the SC decoder and in part to the poor minimum distance of polar codes.
An effective method to fix both of these problems has been to use a concatenation scheme in which a high-rate outer code is used to pre-code the source bits before they go into an inner polar code.
A particularly powerful example of such methods is the CRC-aided SC list decoding (CA-SCL) \cite{tal_list_2011}, whose FER performance is shown in Fig.~\ref{fig:Performance} for the case of $N=128$, $R=1/2$, CRC length 8, and list size 32. In the next section, we consider improving the polar code performance still further by shifting the burden of error correction entirely to an outer code.

\section{Polarization-adjusted convolutional codes}\label{sect:PAC}

\begin{figure}[htb]
\begin{center}
\resizebox{!}{5cm}{
\includegraphics{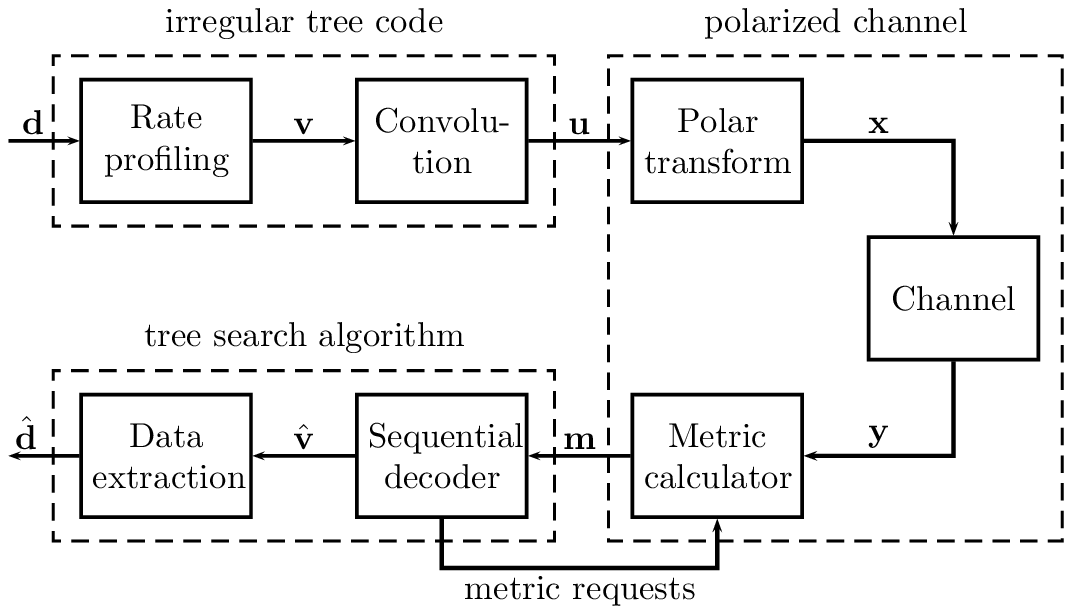}
}
\normalsize
\caption{PAC coding scheme.}
\label{fig:PAC}
\end{center}
\end{figure}

In this section, we consider a new class of codes that we will refer to as {\sl polarization-adjusted convolutional} (PAC) codes. 
The motivating idea for PAC codes is the recognition that 0-1 rate assignments waste the capacities $C(W_i)$ of bit-channels $W_i$ whose inputs are fixed by the rate assignment $R_i=0$.
The capacity loss is especially significant at practical (small to moderate) block-lengths $N$ since polarization takes place relatively slowly.
In order to prevent such capacity loss, we need a scheme that avoids fixing the input of any bit-channel. 
PAC codes achieve this by placing an outer convolutional coding block in front of the polar transform as shown in Fig.~\ref{fig:PAC}.

As with polar codes, the natural block lengths for PAC codes are powers of two, $N=2^n$, $n\ge 1$. The code dimension $K$ can be any integer between 1 and $N$. 
The encoding operation for PAC codes is as follows. A rate-profiling block inserts the source word $\bd$ into a data carrier word $\bv$ in accordance with a data index set $\calA$ so that $\bv_\calA=\bd$ 
and $\bv_{\calA^c}={\mathbf 0}$.
The PAC codeword $\bx$ is obtained from $\bv$ by a one-to-one transformation $\bx =\bv \bT\bP_n$ where $\bT$ is a convolution operation and $\bP_n$ is the polar transform. 
A low-complexity encoding alternative is to compute first $\bu=\bv \bT$ and then $\bx=\bu\bP_n$.

As usual, we characterize the convolution operation by an impulse response $\bc=(c_0,\cdots,c_m)$, where by convention we assume that $c_0\neq 0$ and $c_m\neq 0$. The parameter $m+1$ is called the constraint length of the convolution. The input-output relation for a convolution with a given impulse response $\bc=(c_0,\cdots,c_m)$ is 
$$
u_i = \sum_{j=0}^m c_jv_{i-j}
$$
where it is understood that $v_{i-j}=0$ for $j\ge i$.
The same convolution operation can be represented in matrix form as $\bu=\bv\bT$
where $\bT$ is an upper-triangular Toeplitz matrix,
$$
{\mathbf T} = \begin{bmatrix}
c_0 & c_1 & c_2 & \cdots & c_m & 0 & \cdots & 0\\
0 & c_0   & c_1 &   c_2  &   \cdots  & c_m &      & \vdots\\
0 & 0 & c_0 & c_1 & \ddots & \cdots & c_m& \vdots\\
\vdots & 0 & \ddots & \ddots & \ddots & \ddots & & \vdots\\
\vdots & & \ddots & \ddots & \ddots & \ddots & 0 & \vdots\\
\vdots & & & \ddots & 0 & c_0 & c_1 & c_2\\
\vdots & & & & 0 & 0 & c_0 & c_1\\
0 & \cdots & \cdots  & \cdots & \cdots & 0 & 0 & c_0\\
\end{bmatrix}.
$$

To illustrate the above encoding operation, consider a small example with $N=8$, $K=4$, $\calA=\{4,6,7,8\}$, and $\bc=(1,1,1)$. 
The rate-profiler maps the source word $\bd=(d_1,\ldots,d_4)$ into $\bv=(v_1,\ldots,v_8)$ so that 
$$
{\mathbf v}=(0,0,0,d_1,0,d_2,d_3,d_4).
$$ 
The convolution $\bu=\bv\bT$ generates an output word $\bu$ with $u_1=v_1$, $u_2=v_1+v_2$, and $u_i=v_{i-2}+v_{i-1}+v_{i}$ for $i=3,\ldots,8$.
(This convolution can be implemented as in Fig.~\ref{fig:ConvCode} by taking the upper part of that circuit.)
Encoding is completed by computing the polar transform $\bx=\bu\bP_3$.

\begin{figure}[bht]
\begin{center}
\resizebox{!}{8cm}{
\includegraphics{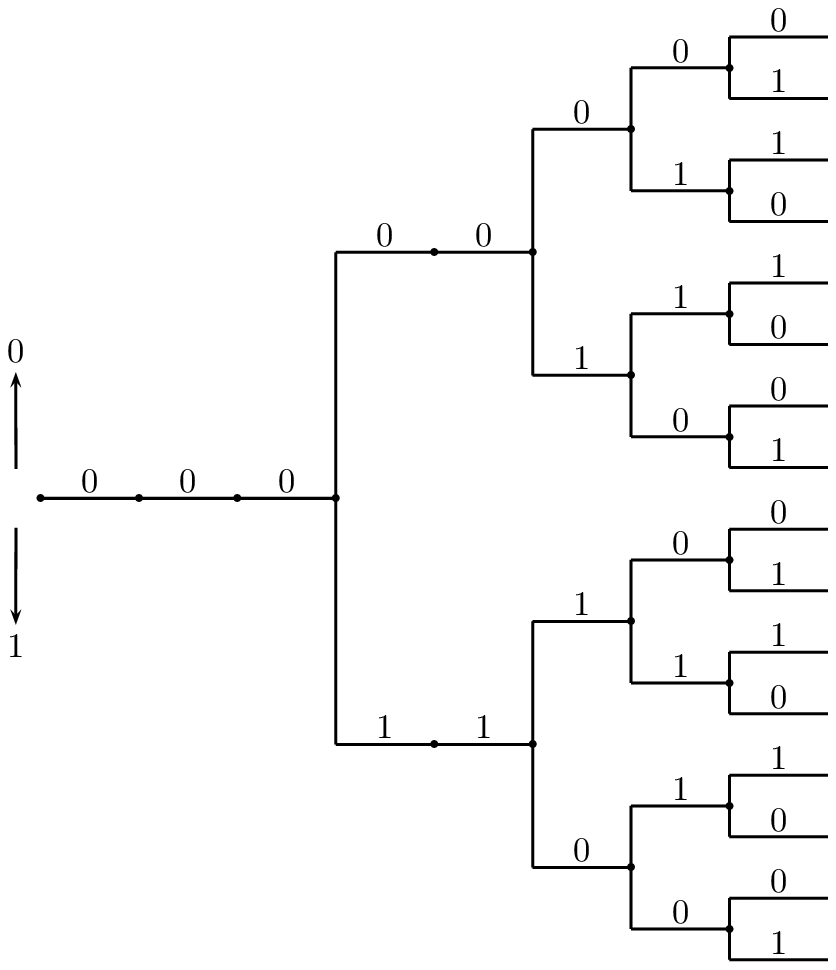}
}
\normalsize
\caption{Irregular tree code example.}
\label{fig:IrrConvCode}
\end{center}
\end{figure}

Unlike ordinary convolutional codes, the convolution operation here generates an irregular tree code due to the constraint $\bv_{\calA^c}={\mathbf 0}$.
Fig.~\ref{fig:IrrConvCode} illustrates the irregular tree code generated by the convolution in the above example.  
The tree in Fig.~\ref{fig:IrrConvCode} branches only at time indices in the set $\calA$, {\sl i.e.,} only when there is a new source bit $d_i$ going into the convolution operation. 
When there is a branching in the tree at some stage $i\in \calA$, by convention, the upper branch corresponds to $v_i=0$ and the lower branch to $v_i=1$. 
Leaf nodes of the tree in Fig.~\ref{fig:IrrConvCode} are in one-to-one correspondence with the convolution input words $\bv$ satisfying the constraint $\bv_{\calA^c}={\mathbf 0}$.
The branches on the path to a leaf node $\bv$ are labeled with the symbols of the convolution output word $\bu=\bv \bT$. 

To summarize, a PAC code is specified by four parameters $(N,K,\calA,\bc)$. 
In simulation studies we observed that the performance of a PAC code is more sensitive to the choice of $\calA$ than to $\bc$. 
As long as the constraint length of the convolution is sufficiently large, choosing $\bc$ at random may be an acceptable design practice.
Finding good design rules for $\calA$ is a research problem.
 
A heuristic method of choosing $\calA$ is to use a {\sl score} function $s:\{1,\ldots,N\}\to {\mathbb R}$ and select $\calA$ as a set of indices $i$ such that $s(i)$ is among the largest $K$ scores in the list 
$s(1),\ldots,s(N)$ (with ties broken arbitrarily). 
Two examples of score functions (inspired by polar codes) are the capacity score function $s(i)=C(W_i)$ and the cutoff rate score function $s(i)=R_0(W_i)$ where 
$\{W_i\}$ are the MLC/MSD bit-channels created by the polar transform $\bP_n$. 
The cutoff rate score function recovers polar codes when $\bT$ is set to the identity transform (corresponding to $\bc=1$). 
A third example of a score function is the Reed-Muller (RM) score function $s(i)=w(i-1)$ where $w(i-1)$ is the number of ones in the binary representation of $i-1$, $0\le i-1\le N-1$. 
For example, $w(12)=2$ since 12 has the binary representation $1100$. 
We refer to this score function as the RM score function since it generates the well-known RM codes \cite{reed_class_1954}, \cite{muller_application_1954} 
when $\bT$ is the identity transform.

We now turn to decoding of PAC codes. For purposes of discussing the decoding operation, it is preferable to 
segment the PAC coding system into three functional blocks as shown by dashed-rectangles in Fig.~\ref{fig:PAC}.
According to this functional segmentation, a source word $\bd$ is inserted into a data carrier $\bv$, the data carrier $\bv$ is encoded into an codeword $\bu$ from an irregular tree code, the codeword $\bu$ is sent over a polarized channel, a sequential decoder is used to generate an estimate $\hat{\bv}$ of $\bv$, and finally, an estimate $\hat{\bd}$ of the source word $\bd$ is extracted from $\hat{\bv}$ by setting $\hat{\bd}=\hat{\bv}_\calA$.

Irregular tree codes can be decoded by tree search heuristics in much the same way as regular tree codes.
A particularly suitable tree search heuristic for PAC codes is sequential decoding, specifically, the Fano decoder \cite{fano_heuristic_1963}.
The Fano decoder tries to identify the correct path in the code tree by using a metric that tends to drift up along the correct path and drift down as soon as a path diverges from the correct path.
The Fano decoder generates metric requests along the path that it is currently exploring and a metric calculator responds by sending back the requested metric values (denoted by ${\mathbf m}$ in Fig.~\ref{fig:PAC}). 
Unlike the usual metric in sequential decoding, the metrics here have to have a time-varying bias so as to maintain the desired drift properties in the face of the irregular nature of the tree code.
In computing the metric, the metric calculator can use a recursive method, as in SC decoding of polar codes.

Fig.~\ref{fig:Performance} presents the result of a computer simulation with a PAC code with $N=128$, $R=1/2$, $\calA$ chosen in accordance with the RM design rule, and $\bc=(1,0,1,1,0,1,1)$.
As seen in the figure, the FER performance of the PAC code in this example comes very close to the dispersion approximation for FER values larger than $10^{-3}$. 
Evidently, the product of the polar transform $\bP_n$ and the convolution transform $\bT$ creates an overall transform $\bG = \bT\bP_n$ that looks sufficiently random to 
achieve a performance near the dispersion approximation. 
When we repeated this simulation experiment with a PAC code designed by the polar coding score function (keeping everything else the same), 
we observed that the performance became worse but the sequential decoder ran significantly faster.
The RM design was the best design we could find in terms of FER performance.

As a heuristic guide to understanding the computational behavior of sequential decoding of a PAC code, we found it useful to associate a {\sl rate profile} to each design rule or equivalently data index set $\calA$.
The rate profile for a data index set $\calA$ is defined as the the sequence of numbers $\{K_i\}_{i=0}^N$ where $K_0=0$ and $K_i$ is the number of elements in $\calA\cap \{1,2,\ldots,i\}$ for $i\ge 1$.
Thus, $K_i$ is the number of source data bits carried in the first $i$ coordinates of the data carrier word $\bv$.
The rate profiles associated with the RM and polar code design rules are shown in Fig.~\ref{fig:Profile} for $N=128$ and $K=64$.
We expect that a design rule whose rate profile stays below the polarized cutoff rate profile at a certain SNR will generate a PAC code that has low complexity under sequential decoding at that SNR.  
In Fig.~\ref{fig:Profile}, both the RM and polar rate profiles lie below the polarized cutoff rate profile, but the polar rate profile leaves a greater safety margin, which may explain the experimental observation 
that the Fano decoder runs faster with the polar code design rule.

\section{Remarks and open problems}

We conclude the note with some complementary remarks about PAC codes and suggestions for further research.

One may view PAC codes as a concatenation scheme with an outer convolutional code and an inner polar code. 
However, PAC codes differ from typical concatenated coding schemes in that the inner code in PAC coding has rate one, so it has no error correction capability. 
It is more appropriate to view the inner polar transform and the metric calculator (mapper and demapper) in PAC coding as a pair of pre- and post-processing devices
around a memoryless channel that provide polarized information to an outer decoder so as to increase the performance of the outer coding system.

In view of the data-processing theorem mentioned in connection with Pinsker's scheme, 
it seems impossible that PAC codes be able to operate at low-complexity at rates above the cutoff rate $R_0(W)$ using only a {\sl single} sequential decoder.
This is true only in part. PAC codes use a convolutional code whose length spans only one use of the polarized channel.
The sequential decoder in PAC coding stops searching for the correct path if a decision error is made after reaching level $N$ in the irregular code tree, {\sl i.e.},
after a single use of the polarized channel.
The $R_0(W)$ bound on sequential decoding would hold if a convolutional code were used that extended over multiple uses of the polarized channel.
A better understanding of the computational complexity of the sequential decoder in PAC coding is an open problem. 

As stated above, the performance and complexity of PAC codes are yet to be studied rigorously. 
It is clear that in general PAC codes can achieve channel capacity since they contain polar codes as a special case. 
The main question is to characterize the best attainable performance by PAC codes over variation of the data index set $\calA$ and the convolution impulse response $\bc$.

The fact that PAC codes perform well under the RM design rule suggests that, unlike polar codes, PAC codes are robust against channel parameter variations and modeling errors.
It is of interest to investigate if PAC codes have \emph{universal} design rules so that a given PAC code performs well uniformly over the class of all BMCs with a given capacity.
In particular, it is of interest to check if the RM design rule (together with a suitably chosen convolution impulse response $\bc$) is universal in this sense.

A disadvantage of the sequential decoding method is its variable complexity. It is of interest to study fixed-complexity search heuristics for decoding PAC codes.  
One possibility is to use a breadth-first search heuristic, such as a Viterbi decoder.
However, a Viterbi decoder that tracks only the state of the convolutional encoder will be suboptimal since PAC codes incorporate a polarized channel that, too, has a state.
In fact, the number of states of the polarized channel is the same as the number of possible words $\bu$ at the input of the polarized channel, namely, $2^{NR}$ for a PAC code of length $N$ and rate $R$.
There is clearly need for a sub-optimal breadth-first search heuristic that tracks only a subset of all possible states. 
One option that may be considered here is list Viterbi decoding \cite{seshadri_list_1994} which is a method that has proven effective for searching large state spaces.
For some other alternatives of forward pruning methods in breadth-first search, such as beam search, we refer to \cite[pp.~174-175]{russell_artificial_2009}.

In linear algebra, lower-upper decomposition (LUD) is a method for solving systems of linear equations.
PAC coding may be regarded as one form of upper-lower decomposition (ULD) of a code generator matrix $G$ for purposes of solving a redundant set of linear equations when the equations are corrupted by noise. 
One may investigate if there are other decompositions in linear algebra for synthesizing generator matrices that yield powerful codes with low-complexity encoding and decoding.

\end{document}